\documentclass[]{hhr_arxiv}

\title{BINARY SELF-SELECTIVE VOTING RULES}

\newcommand{\myauthorA}{\textsc{H\'{e}ctor Hermida-Rivera}}%
\newcommand{\myemailA}{\href{mailto:hermida.rivera.hector@gtk.bme.hu}{\texttt{hermida.rivera.hector@gtk.bme.hu}}}%
\newcommand{\myaffiliationA}{\linespread{1.2}\selectfont\emph{Quantitative Social and Management Sciences Research Centre, Faculty of Economic and Social Sciences, Budapest University of Technology and Economics, Műegyetem rkp. 3, Budapest H‑1111, Hungary.}
}%

\newcommand{\myauthorB}{\textsc{Toygar T. Kerman}}%
\newcommand{\myemailB}{\href{mailto:toygar.kerman@uni-corvinus.hu}{\texttt{toygar.kerman@uni-corvinus.hu}}}%
\newcommand{\myaffiliationB}{\linespread{1.2}\selectfont\emph{Institute of Economics, Corvinus University of Budapest, F\H{o}v\'{a}m t\'{e}r 8, Budapest, 1093, Hungary.}\\
}%

\begin{document}

\titletwo


\begin{abstract}
This paper introduces a novel binary stability property for voting rules---called \hyperref[b]{binary self-selectivity}---by which a society considering whether to replace its voting rule using itself in pairwise elections will choose not to do so. In \Cref{th}, we show that a \hyperref[n]{neutral} voting rule is \hyperref[b]{binary self-selective} if and only if it is \hyperref[s]{universally self-selective}. We then use this equivalence to show, in \Cref{col1}, that under the unrestricted strict preference domain, a \hyperref[u]{unanimous} and \hyperref[n]{neutral} voting rule is \hyperref[b]{binary self-selective} if and only if it is \hyperref[d]{dictatorial}. In \Cref{col2,col3}, we show that whenever there is a strong Condorcet winner; a \hyperref[u]{unanimous}, \hyperref[n]{neutral} and \hyperref[a]{anonymous} voting rule is \hyperref[b]{binary self-selective} (or \hyperref[s]{universally self-selective}) if and only if it is the \hyperref[con]{Condorcet voting rule}.
\end{abstract}


\jelcodes{D71, D82.}

\wordcount{5,253}


\section{Introduction}\label{sec.int}

\textsc{Historically}, virtually all countries that have decided whether to change their voting rule have done so in a binary fashion. For example, in 1918, Switzerland held a referendum to decide between keeping a three-round majoritarian system or introducing proportional representation, which was the option chosen by 66.8\% of the popular vote. In 1993, New Zealand held a referendum in which voters chose to replace the first-past-the-post method for a mixed-member proportional representation method with a 53.6\% of the popular vote. More recently, in 2011, the United Kingdom choose to keep the first-past-the-post method in a referendum against the instant run-off method with a 67.9\% of the popular vote. In all these three referenda---and in most such referenda across the world---the choice was binary: namely, between keeping or replacing the country's existing voting rule in a head-to-head contest.

In this paper, we introduce a novel \hyperref[b]{binary self-selectivity} axiom that draws its inspiration from \textcite{koray_00}, who proposed a consistency property---called \hyperref[s]{universal self-selectivity}---by which a \hyperref[n]{neutral} voting rule used by society to select one alternative among many should also select itself when used to select a voting rule among any finite set of \hyperref[n]{neutral} voting rules it belongs to. \citeauthor{koray_00}'s (\citeyear{koray_00}, Theorem 2, p. 990) central result states that under the unrestricted domain of strict preferences, a \hyperref[n]{neutral} and \hyperref[u]{unanimous} voting rule is \hyperref[s]{universally self-selective} if and only if it is \hyperref[d]{dictatorial}.

In his paper, \textcite{koray_00} assumes that a voting rule must select itself among any finite set of \hyperref[n]{neutral} voting rules it belongs to. However, most societies decide on their voting rule by voting in favor or against some alternative voting rule: that is, in binary elections. One may then naturally wonder whether \citeauthor{koray_00}'s (\citeyear{koray_00}, Theorem 2, p. 990) result holds when demanding that a \hyperref[n]{neutral} voting rule selects itself only from sets of voting rules containing itself and \emph{just one} other \hyperref[n]{neutral} voting rule. This paper addresses this question, as we modify \citeauthor{koray_00}'s \citeyearpar{koray_00} \hyperref[s]{universal self-selectivity} by assuming that society chooses whether to replace its voting rule in a binary fashion.

In \Cref{th}, we show that replacing \hyperref[s]{universal self-selectivity} for \hyperref[b]{binary self-selectivity} makes no difference, for a \hyperref[n]{neutral} voting rule is \hyperref[b]{binary self-selective} if and only if it is \hyperref[s]{universally self-selective}. Hence, in \Cref{col1}, we show that \hyperref[d]{dictatorships} are the only \hyperref[u]{unanimous}, \hyperref[n]{neutral} and \hyperref[b]{binary self-selective} voting rules. Further, we also show in \Cref{col2,col3} that under the restricted domain of strict preferences with a strong Condorcet winner; a \hyperref[u]{unanimous}, \hyperref[n]{neutral} and \hyperref[a]{anonymous} voting rule is \hyperref[b]{binary self-selective} (or \hyperref[s]{universally self-selective}) if and only if it is the \hyperref[con]{Condorcet voting rule} (i.e., the majority voting rule). All in all, \hyperref[b]{binary self-selectivity} is a more natural modeling choice than \hyperref[s]{universal self-selectivity}, but it is not a technical requirement to obtain our results. 

Our results reinforce the long-standing idea that the \hyperref[con]{Condorcet voting rule} is special and the most desirable voting rule whenever it produces a unique winner. Notably, \textcite[Theorem 2, p. 965]{dasguptamaskin_08} show that the majority voting rule satisfies five highly desirable axioms---\hyperref[p]{Pareto}, \hyperref[a]{anonymity}, \hyperref[n]{neutrality}, \hyperref[iia]{independence of irrelevant alternatives} and \emph{decisiveness}---over a larger class of preference domains than any other voting rule. In a related paper, \textcite[Theorem 2, p. 467]{dasguptamaskin_20} show that \emph{group strategy-proofness}, \hyperref[p]{Pareto}, \hyperref[a]{anonymity}, \hyperref[n]{neutrality}, \hyperref[iia]{independence of irrelevant alternatives} and \emph{decisiveness} uniquely characterize the majority voting rule on any domain of preferences for which there exists a voting rule satisfying these axioms (namely, on any restricted domain without Condorcet cycles). In our paper, we add to this body of work by showing that whenever a strong Condorcet winner exists, the majority voting rule is the only democratic and stable voting rule.

\citeauthor{koray_00}'s \citeyearpar{koray_00} analysis has been extended in a few directions. Notably, several authors have successfully escaped his impossibility result. For example, \textcite{unel_99} restricts the domain of voting rules to single-peaked preference profiles, and \textcite{korayslinko_08,kermankoray_24} restrict the set of rival functions society shall choose from. However, some authors have shown that certain restricting assumptions are not enough to escape \citeauthor{koray_00}'s \citeyearpar{koray_00} impossibility result. For example, \textcite{korayunel_03} restrict the set of rival functions to \emph{tops-only} ones; whereas \textcite{koraysenocak_24} introduce a weaker notion of stability called \emph{selection-closedness}. Finally, some authors have explored other related issues, such as the manipulability of \hyperref[s]{universally self-selective} voting rules \parencite{diss_15}, the \hyperref[s]{universal self-selectivity} of triplets of voting rules \parencite{dissmerlin_10,disslouichimerlinsmaoui_12}, the properties of \emph{dynamically stable} sets of constitutions \parencite{houy_07}, or the stability of the French and (failed) European constitutions \parencite{houy_06}.

Other authors have also analyzed closely related problems, using simple games \parencite{vonneumannmorgenstern_44} as voting rules. Notably, \textcite{barberajackson_04} introduce a model of binary ex-ante self-stability, inspired by \textcite{rae_69}, in which every voter's preference is modeled through some probability distribution indicating the likelihood this voter will favor reform. Their seminal contribution has been extended in several directions. For example, \textcite{azrielikim_16} expand their model to weighted majority voting rules; \textcite{jeongkim_23} present the concepts of interim and ex-post self-stability; \textcite{jeongkim_24} bridge the gap between ex post and ex ante stability; \textcite{kulttimiettinen_07} characterize the \textcite{vonneumannmorgenstern_44} stable set of majority voting rules; \textcite{coelho_05} characterizes the set of voting rules satisfying the maximin criterion; \textcite{kulttimiettinen_09} consider constitutions with arbitrary many voting rules; \textcite{lagunoff_09} introduces a dynamic model of stability; and \textcite{hermidarivera_24} introduces the notion of minimal stability.

The rest of the paper is organized as follows: \Cref{sec.env} defines the environment, \Cref{sec.ax} introduces the axioms, and \Cref{sec.res} contains the results.

\section{Environment}\label{sec.env}

The \emph{environment} is a $4$-tuple $(N,A_\tau,\mathcal{R}(A_\tau),\sigma)$. Let $N=\{1,\dots,n\}$ be a \emph{finite voter set}, where $n\geqslant2$. Let $A_\tau=\{a_1,\dots,a_\tau\}$ be a \emph{finite alternative set}, where $\tau\in\mathbb{N}=\{1,2,\dots\}$. Let
\begin{gather}
    \mathcal{R}(A_\tau)=\bigl\{R\mid R=(R_i)_{i\in N}:R_i\text{ is \emph{voter $i$'s weak order} over }A_\tau\bigr\}
\end{gather}
be the \emph{set of all weak preference profiles over some alternative set} whose \emph{strict part} is $\mathcal{P}(A_\tau)=\{P\mid P=(P_i)_{i\in N}\}$: namely, for all alternatives $x,y\in A_\tau$ and all voters $i\in N$, we have $x R_iy$ if and only if voter $i$ \emph{weakly prefers} $x$ over $y$, and $x P_iy$ if and only if voter $i$ \emph{strictly prefers} $x$ over $y$. Finally, given any $\tau\in\mathbb{N}$ and any \emph{set of admissible preference profiles} $\mathcal{A}(A_\tau)\subseteq\mathcal{P}(A_\tau)$, let
\begin{gather}
    \sigma:\bigcup_{\tau\in\mathbb{N}}\mathcal{A}(A_\tau)\to\bigcup_{\tau\in\mathbb{N}}A_\tau
\end{gather}
be a \emph{voting rule} so that $\sigma(P)\in A_\tau$ for all $\tau\in\mathbb{N}$ and all profiles $P\in\mathcal{A}(A_\tau)$.

\section{Axioms}\label{sec.ax}

On the one hand, the \hyperref[u]{unanimity}, \hyperref[d]{dictatorship}, \hyperref[a]{anonymity} and \hyperref[n]{neutrality} axioms are well-known axioms that barely need an introduction; on the other hand, the \hyperref[b]{binary self-selectivity} axiom is a natural adaptation of \citeauthor{koray_00}'s (\citeyear{koray_00}, p. 985) \hyperref[s]{universal self-selectivity} to binary environments in which society chooses whether to replace its existing voting rule in head-to-head contests with all other voting rules.

Let $\Sigma=\{\sigma\mid\sigma\text{ is a voting rule}\}$.

Given any $\tau\in\mathbb{N}$ and any profile $P\in\mathcal{P}(A_\tau)$, let $t_i(P)$ be \emph{voter $i$'s top-ranked alternative at profile $P$}.

\begin{axiom}[Unanimity]\label{u}
    A voting rule $\sigma$ is \emph{unanimous} if and only if it always selects the unanimously top-ranked alternative whenever there is one. Formally, if and only if
\begin{gather}
    (\forall\tau\in\mathbb{N})(\forall P\in\mathcal{A}(A_\tau))(\forall x\in A_\tau)[((\forall i\in N)(t_i(P)=x))\Rightarrow(\sigma(P)=x)]
\end{gather}
\end{axiom}

Let $\Sigma_u=\{\sigma\in\Sigma\mid\sigma\text{ is \hyperref[u]{unanimous}}\}$.

\begin{axiom}[Dictatorship]\label{d}
    A voting rule $\sigma$ is \emph{dictatorial} if and only if it always selects the same voter's top-ranked alternative. Formally, if and only if
\begin{gather}\label{eq.d}
    (\exists i\in N)[(\forall\tau\in\mathbb{N})(\forall P\in\mathcal{A}(A_\tau))(\sigma(P)=t_i(P))]
\end{gather}
\end{axiom}

Given any voter $i\in N$, let $\delta_i$ be his \hyperref[d]{dictatorship}: namely, $\delta_i(P)=t_i(P)$ for all $\tau\in\mathbb{N}$ and all profiles $P\in\mathcal{A}(A_\tau)$.

Let $\Delta=\{\sigma\in\Sigma\mid\sigma\text{ is \hyperref[d]{dictatorial}}\}$.

Let $\Pi=\{\pi:N\to N\mid\pi\text{ is bijective}\}$ be the \emph{set of all permutations of the voter set}. Then, given any $\tau\in\mathbb{N}$, any profile $P\in\mathcal{P}(A_\tau)$ and any permutation $\pi\in\Pi$, let the profile $\pi P=(P_{\pi(i)})_{i\in N}\in\mathcal{P}(A_\tau)$ satisfy, for all voters $i\in N$ and all alternatives $x,y\in A_\tau$, $x\pi P_iy$ if and only if $xP_{\pi(i)}y$.

\begin{axiom}[Anonymity]\label{a}
    A voting rule $\sigma$ is \emph{anonymous} if and only if it is independent of voters' identities. Formally, if and only if
\begin{gather}\label{eq.a}
    (\forall\tau\in\mathbb{N})(\forall P\in\mathcal{A}(A_\tau))(\forall\pi\in\Pi)[\sigma(P)=\sigma(\pi P)]
\end{gather}
\end{axiom}

Let $\Sigma_a=\{\sigma\in\Sigma\mid\sigma\text{ is \hyperref[a]{anonymous}}\}$.

Given any $\tau\in\mathbb{N}$, let $M_\tau=\{\mu:A_\tau\to A_\tau\mid\mu\text{ is bijective}\}$ be the \emph{set of all permutations of the alternative set}. Then, given any profile $P\in\mathcal{P}(A_\tau)$ and any permutation $\mu\in M_\tau$, let the profile $\mu P\in\mathcal{P}(A_\tau)$ satisfy, for all voters $i\in N$ and all alternatives $x,y\in A_\tau$, $\mu(x)\mu P_i\mu(y)$ if and only if $x P_iy$.

\begin{axiom}[Neutrality]\label{n}
    A voting rule $\sigma$ is \emph{neutral} if and only if it is independent of alternatives' labels. Formally, if and only if
\begin{gather}\label{eq.n}
    (\forall\tau\in\mathbb{N})(\forall P\in\mathcal{A}(A_\tau))(\forall\mu\in M_\tau)[\mu(\sigma(P))=\sigma(\mu P)]
\end{gather}
\end{axiom}

Let $\Sigma_n=\{\sigma\in\Sigma\mid\sigma\text{ is \hyperref[n]{neutral}}\}$.

The domain of \hyperref[n]{neutral} voting rules can be easily extended to strict preference profiles on any finite non-empty set. Given any set $X$ with cardinality $\tau$, let $B(X)=\{\beta:X\to A_\tau\mid\beta\text{ is bijective}\}$ be the \emph{set of all bijections} from $X$ onto $A_\tau$. Given any profile $P\in\mathcal{P}(X)$ and any bijection $\beta\in B(X)$, let the profile $\beta P\in\mathcal{P}(A_\tau)$ satisfy, for all voters $i\in N$ and all elements $a,b\in X$, $aP_ib$ if and only if $\beta(a)\beta P_i\beta(b)$. Then, given any two bijections $\beta,\tilde{\beta}\in B(X)$, any profile $P\in\mathcal{P}(X)$ and any \hyperref[n]{neutral} voting rule $\sigma\in \Sigma_n$, it follows that $\beta^{-1}(\sigma(\beta P))=\tilde{\beta}^{-1}(\sigma(\tilde{\beta}P))$ as long as $\beta P,\tilde{\beta}P\in\mathcal{A}(A_\tau)$. Hence, set $\sigma(P)=\beta^{-1}(\sigma(\beta P))$ for all profiles $P\in\mathcal{P}(X)$ and some bijection $\beta\in B(X)$.

In order to introduce the \hyperref[b]{binary self-selectivity} axiom, we now specify voters' preferences over voting rules. We do so by taking the consequentialist approach, by which preferences over voting rules are derived from the alternatives they choose; rather than the non-consequentialist approach, by which preferences over voting rules are defined independently of the alternatives.

Given any $\tau\in\mathbb{N}$, any profile $P\in\mathcal{P}(A_\tau)$ naturally induces a preference profile over any finite set $T\subsetneq\Sigma_n$. Given any $\tau\in\mathbb{N}$ and any profile $P\in\mathcal{P}(A_\tau)$, let the weak profile $\boldsymbol{R}^{T\!P}\in\mathcal{R}(T)$ satisfy, for all voters $i\in N$ and all voting rules $\sigma,\sigma'\in T$, $\sigma \boldsymbol{R}_i^{T\!P}\sigma'$ if and only if $\sigma(P)P_i\sigma'(P)$.

Given any finite set $T\subsetneq\Sigma_n$ and any profile $P\in\mathcal{P}(A_\tau)$, a profile $\boldsymbol{P}\in\mathcal{P}(T)$ is \emph{compatible} with the weak profile $\boldsymbol{R}^{T\!P}$ if and only if for all voting rules $\sigma,\sigma'\in T$ and all voters $i\in N$, we have that $\sigma \boldsymbol{P}_i\sigma'$ implies $\sigma \boldsymbol{R}_i^{T\!P}\sigma'$ (i.e., if and only if $\boldsymbol{P}$ can be obtained from $\boldsymbol{R}^{T\!P}$ by linearly ordering the elements in each indifference class). Then, given any $\tau\in\mathbb{N}$, any finite set $T\subsetneq \Sigma_n$ and any profile $P\in\mathcal{A}(A_\tau)$, let $\mathcal{A}(T,P)=\{\boldsymbol{P}\in\mathcal{A}(T)\mid\boldsymbol{P}\text{ is compatible with }\boldsymbol{R}^{T\!P}\}$.

\begin{axiom}[Binary self-selectivity]\label{b}
    A \hyperref[n]{neutral} voting rule $\sigma\in\Sigma_n$ is \emph{binary self-selective} if and only if given any other \hyperref[n]{neutral} voting rule and any preference profile over the alternatives, there exists a compatible preference profile at which it selects itself. Formally, if and only if
\begin{gather}
    (\forall\tau\in\mathbb{N})(\forall P\in\mathcal{A}(A_\tau))(\forall\sigma'\in\Sigma_n\backslash\{\sigma\})[(\exists\boldsymbol{P}\in\mathcal{A}(\{\sigma,\sigma'\},P))(\sigma(\boldsymbol{P})=\sigma)]
\end{gather}
\end{axiom}

Let $\Sigma_b=\{\sigma\in \Sigma\mid\sigma\text{ is \hyperref[b]{binary self-selective}}\}$.

We now introduce \citeauthor{koray_00}'s (\citeyear{koray_00}) \hyperref[s]{universal self-selectivity} axiom, but on any domain (rather than only on the unrestricted domain of strict preferences). Given any \hyperref[n]{neutral} voting rule $\sigma\in\Sigma_n$, let $\mathcal{T}(\sigma)=\{T\subsetneq\Sigma_n\mid(\sigma\in T)\wedge(|T|<\infty)\}$ be the \emph{collection of all finite sets of \hyperref[n]{neutral} voting rules containing the voting rule $\sigma$}.

\begin{axiom}[Universal self-selectivity]\label{s}
    A \hyperref[n]{neutral} voting rule $\sigma\in\Sigma_n$ is \emph{universally self-selective} if and only if given any finite set of \hyperref[n]{neutral} voting rules it belongs to and any preference profile over the alternatives, there exists a compatible preference profile at which it selects itself. Formally, if and only if
\begin{gather}
    (\forall\tau\in\mathbb{N})(\forall P\in\mathcal{A}(A_\tau))(\forall T\in\mathcal{T}(\sigma))[(\exists\boldsymbol{P}\in\mathcal{A}(T,P))(\sigma(\boldsymbol{P})=\sigma)]
\end{gather}
\end{axiom}

Let $\Sigma_s=\{\sigma\in \Sigma\mid\sigma\text{ is \hyperref[s]{universally self-selective}}\}$.

\begin{exampleb}
    Consider a society with five voters $N=\{1,\dots,5\}$ and four alternatives $A_4=\{x,y,z,w\}$. Let $p$ be the \emph{plurality rule} that selects the alternative top-ranked by the largest number of voters if there is only one, and voter $1$'s top-ranked alternative otherwise; and let $b$ be the \emph{Borda rule} that selects the alternative with the highest Borda count with scoring vector $(\tau,\tau-1,\dots,1)$ if there is only one, and voter $4$'s top-ranked alternative otherwise.
\begin{table}[!ht]
\centering
\caption{Strict preference profile $P$}
\label{preference}
\medskip
\renewcommand{\arraystretch}{1.15}
\begin{tabularx}{\textwidth}{cXcXcXcXc}\hline\hline
 $P_1$ & & $P_2$ & & $P_3$ & & $P_4$ & & $P_5$ \\\hline 
 $x$ & & $x$ & & $x$ & & $y$ & & $y$ \\ 
 $y$ & & $y$ & & $y$ & & $z$ & & $z$ \\
 $z$ & & $z$ & & $z$ & & $w$ & & $w$ \\
 $w$ & & $w$ & & $w$ & & $x$ & & $x$ \\\hline\hline
\end{tabularx}
\end{table}

Suppose now that society's preferences are given by the profile $P=(P_i)_{i\in N}\in\mathcal{P}(A_4)$ in \Cref{preference}, and that its choice is between the plurality and the Borda rules, so that $T=\{p,b\}$. Then, $p(P)=x$ and $b(P)=y$. Thus, the unique compatible profile $\boldsymbol{P}\in\mathcal{P}(T,P)$ is given by \Cref{preference1}. Then, $p(\boldsymbol{P})=b(\boldsymbol{P})=p$. Hence, the Borda rule is not \hyperref[b]{binary self-selective}.
\begin{table}[!ht]
\centering
\caption{Preference profile $\boldsymbol{P}$}
\label{preference1}
\medskip
\renewcommand{\arraystretch}{1.15}
\begin{tabularx}{\textwidth}{cXcXcXcXc}\hline\hline
 $\boldsymbol{P}_1$ & & $\boldsymbol{P}_2$ & & $\boldsymbol{P}_3$ & & $\boldsymbol{P}_4$ & & $\boldsymbol{P}_5$ \\\hline 
 $p$ & & $p$ & & $p$ & & $b$ & & $b$ \\ 
 $b$ & & $b$ & & $b$ & & $p$ & & $p$ \\\hline\hline
\end{tabularx}
\end{table}

Suppose now that society's preferences are given by the profile $P'=(P'_i)_{i\in N}\in\mathcal{P}(A_4)$ in \Cref{preference'}, and that its choice is again between the plurality and the Borda rules, so that $T=\{p,b\}$. Then, $p(P')=x$ and $b(P')=y$. Thus, the unique compatible profile $\boldsymbol{P}'\in\mathcal{P}(T,P')$ is given by \Cref{preference1'}. Then, $p(\boldsymbol{P}')=b(\boldsymbol{P}')=b$. Hence, the plurality rule is not \hyperref[b]{binary self-selective}.
\begin{table}[!ht]
\centering
\caption{Strict preference profile $P'$}
\label{preference'}
\medskip
\renewcommand{\arraystretch}{1.15}
\begin{tabularx}{\textwidth}{cXcXcXcXc}\hline\hline
 $P'_1$ & & $P'_2$ & & $P'_3$ & & $P'_4$ & & $P'_5$ \\\hline 
 $x$ & & $x$ & & $y$ & & $z$ & & $w$ \\ 
 $y$ & & $y$ & & $w$ & & $y$ & & $y$ \\
 $w$ & & $z$ & & $z$ & & $w$ & & $z$ \\
 $z$ & & $w$ & & $x$ & & $x$ & & $x$ \\\hline\hline
\end{tabularx}
\end{table}
\begin{table}[!ht]
\centering
\caption{Preference profile $\boldsymbol{P}'$}
\label{preference1'}
\medskip
\renewcommand{\arraystretch}{1.15}
\begin{tabularx}{\textwidth}{cXcXcXcXc}\hline\hline
 $\boldsymbol{P}'_1$ & & $\boldsymbol{P}'_2$ & & $\boldsymbol{P}'_3$ & & $\boldsymbol{P}'_4$ & & $\boldsymbol{P}'_5$ \\\hline 
 $p$ & & $p$ & & $b$ & & $b$ & & $b$ \\ 
 $b$ & & $b$ & & $p$ & & $p$ & & $p$ \\\hline\hline
\end{tabularx}
\end{table}
\end{exampleb}

\section{Results}\label{sec.res}

\begin{theorem}\label{th}
    Given any set of admissible preference profiles, a \hyperref[n]{neutral} voting rule is \hyperref[b]{binary self-selective} if and only if it is \hyperref[s]{universally self-selective}. Formally,
\begin{gather}
    [(\forall\tau\in\mathbb{N})(\mathcal{A}(A_\tau)\subseteq\mathcal{P}(A_\tau))]\Rightarrow[(\sigma\in\Sigma_n)\Rightarrow((\sigma\in\Sigma_b)\iff(\sigma\in\Sigma_s))]
\end{gather}
\end{theorem}

\begin{proof}
    Consider any $\tau\in\mathbb{N}$ and any set of admissible preference profiles $\mathcal{A}(A_\tau)\subseteq\mathcal{P}(A_\tau)$. Then, there are two statements to show:
\begin{enumerate}
    \item $(\sigma\in\Sigma_n)\Rightarrow((\sigma\in\Sigma_s)\Rightarrow(\sigma\in\Sigma_b))$, 
    \item $(\sigma\in\Sigma_n)\Rightarrow((\sigma\in\Sigma_b)\Rightarrow(\sigma\in\Sigma_s))$.
\end{enumerate}

\begin{statement}\label{s1}
    $(\sigma\in\Sigma_n)\Rightarrow((\sigma\in\Sigma_s)\Rightarrow(\sigma\in\Sigma_b))$.
\end{statement}

    The proof is direct. Consider any \hyperref[n]{neutral} and \hyperref[s]{universally self-selective} voting rule $\sigma\in\Sigma_n\cap\Sigma_s$ as well as any set $T=\{\sigma,\sigma'\}$, where $\sigma'\in\Sigma_n\backslash\{\sigma\}$. By the \hyperref[s]{universal self-selectivity} axiom, $\sigma(\boldsymbol{P})=\sigma$ for some compatible preference profile $\boldsymbol{P}\in\mathcal{A}(T,P)$. But then, $\sigma\in\Sigma_b$.

\begin{statement}\label{s2}
    $(\sigma\in\Sigma_n)\Rightarrow((\sigma\in\Sigma_b)\Rightarrow(\sigma\in\Sigma_s))$.
\end{statement}

    We now introduce one new axiom that will be useful to prove \Cref{s2}.

\begin{axiom}[Independence of irrelevant alternatives]\label{iia}
    A voting rule $\sigma$ is \emph{independent of irrelevant alternatives} if and only if removing a set of losing alternatives does not change the winning alternative. Formally, if and only if
\begin{gather}
    (\forall\tau\in\mathbb{N})(\forall P\in\mathcal{A}(A_\tau))(\forall B\subseteq A_\tau\backslash\{\sigma(P)\})(\sigma(P)=\sigma(P|_{A_\tau\backslash B}))
\end{gather}
\end{axiom}

    Let $\Sigma_i=\{\sigma\in\Sigma\mid\sigma\text{ is \hyperref[iia]{independent of irrelevant alternatives}}\}$.

    Now, the proof of \Cref{s2} follows from these two claims:
\begin{enumerate}[label=2.\arabic*.]
    \item $\Sigma_n\cap\Sigma_b\subseteq\Sigma_i$,
    \item $\Sigma_n\cap\Sigma_b\subseteq\Sigma_s$.
\end{enumerate}

\begin{claim}\label{c21}
    $\Sigma_n\cap\Sigma_b\subseteq\Sigma_i$.
\end{claim}

    The proof is by contradiction. Consider any \hyperref[n]{neutral}, \hyperref[b]{binary self-selective} and \emph{not} \hyperref[iia]{independent of irrelevant alternatives} voting rule $\sigma\in(\Sigma_n\cap\Sigma_b)\backslash\Sigma_i$, any profile $P\in\mathcal{A}(A_\tau)$, and any set $B\subseteq A_\tau\backslash\{\sigma(P)\}$. If $A_\tau\backslash B=\{\sigma(P)\}$ or $B=\emptyset$, it trivially follows that $\sigma(P)=\sigma(P|_{A_\tau\backslash B})$. Hence, consider any alternative $y\in A_\tau\backslash\{\sigma(P)\}$ and let $A_\tau\backslash B=\{\sigma(P),y\}$ as well as $B\neq\emptyset$.
    
    Consider any \hyperref[n]{neutral} voting rule $\sigma'\in\Sigma_n$ satisfying $\sigma'(P)=y$. Let $T=\{\sigma,\sigma'\}$ and $T(P)=\{\sigma(P),\sigma'(P)\}$. Consequently, $\mathcal{P}(T,P)=\{\boldsymbol{P}\}$; and by the \hyperref[b]{binary self-selectivity} axiom, $\mathcal{P}(T,P)\cap\mathcal{A}(T,P)\neq\emptyset$. Hence, $\mathcal{A}(T,P)=\{\boldsymbol{P}\}$. Let $\beta:T(P)\to T$ satisfy $\beta(\sigma(P))=\sigma$ and $\beta(y)=\sigma'$. Then, $\beta P|_{\{\sigma(P),y\}}=\boldsymbol{P}$. Thus, the \hyperref[n]{neutrality} and \hyperref[b]{binary self-selectivity} axioms together imply that $\sigma=\sigma(\boldsymbol{P})=\sigma(\beta P|_{\{\sigma(P),y\}})=\beta(\sigma(P|_{\{\sigma(P),y\}}))$, further implying that $\sigma(P|_{\{\sigma(P),y\}})=\beta^{-1}(\sigma)=\sigma(P)$.

    Consider any $\tau'\in\mathbb{N}$ satisfying $\tau'>\tau$, and any profile $P'\in\mathcal{A}(A_{\tau'})$ satisfying $P'|_{A_\tau}=P$ and $\sigma(P')\in A_\tau$. Since $\sigma\notin\Sigma_i$, suppose that $\sigma(P')\neq\sigma(P)$. Consider any alternative $z\in A_{\tau'}\backslash\{\sigma(P')\}$ and let $C\subseteq A_{\tau'}$ satisfy $A_{\tau'}\backslash C=\{\sigma(P'),z\}$. Consider any \hyperref[n]{neutral} voting rule $\sigma''\in\Sigma_n$ satisfying $\sigma''(P')=z$. Let $T'=\{\sigma,\sigma''\}$ and $T'(P')=\{\sigma(P'),\sigma''(P')\}$. Consequently, $\mathcal{P}(T',P')=\{\boldsymbol{P}'\}$; and by the \hyperref[b]{binary self-selectivity} axiom, $\mathcal{P}(T',P')\cap\mathcal{A}(T',P')\neq\emptyset$. Then, $\mathcal{A}(T',P')=\{\boldsymbol{P}'\}$. Let $\tilde{\beta}:T'(P')\to T'$ satisfy $\tilde{\beta}(\sigma(P'))=\sigma$ and $\tilde{\beta}(z)=\sigma''$. Then, $\tilde{\beta} P'|_{\{\sigma(P'),z\}}=\boldsymbol{P}'$. Thus, the \hyperref[n]{neutrality} and \hyperref[b]{binary self-selectivity} axioms together imply that $\sigma=\sigma(\boldsymbol{P}')=\sigma(\tilde{\beta} P'|_{\{\sigma(P'),z\}})=\tilde{\beta}(\sigma(P'|_{\{\sigma(P'),z\}}))$, further implying that $\sigma(P'|_{\{\sigma(P'),z\}})=\tilde{\beta}^{-1}(\sigma)=\sigma(P')$.
    
    Since $P'|_{A_\tau}=P$ and $\{\sigma(P),\sigma(P')\}\subseteq A_\tau$, it follows that $P|_{\{\sigma(P),\sigma(P')\}}=P'|_{\{\sigma(P),\sigma(P')\}}$. Therefore, $\sigma(P)=\sigma(P|_{\{\sigma(P),\sigma(P')\}})=\sigma(P'|_{\{\sigma(P),\sigma(P')\}})=\sigma(P')$, which contradicts the assumption that $\sigma(P')\neq\sigma(P)$. Hence, $\sigma\in\Sigma_i$. Thus, $\Sigma_n\cap\Sigma_b\subseteq\Sigma_i$.

\begin{claim}\label{c22}
    $\Sigma_n\cap\Sigma_b\subseteq\Sigma_s$.
\end{claim}

    The proof is by contradiction. Consider any \hyperref[n]{neutral}, \hyperref[b]{binary self-selective} and non-\hyperref[s]{universally self-selective} voting rule $\sigma\in(\Sigma_n\cap\Sigma_b)\backslash\Sigma_s$. Then, there exists some set $T\in\mathcal{T}(\sigma)$ and some admissible profile $P\in\mathcal{A}(A_\tau)$ such that $\sigma(\boldsymbol{P})\neq\sigma$ for all compatible profiles $\boldsymbol{P}\in\mathcal{A}(T,P)$. Fix any such set $T\in\mathcal{T}(\sigma)$ and any such admissible profile $P\in\mathcal{A}(A_\tau)$. By \Cref{c21}, $\Sigma_n\cap\Sigma_b\subseteq\Sigma_i$. Therefore, $\sigma\in\Sigma_i$. Now, by the \hyperref[iia]{independence of irrelevant alternatives} axiom, $\sigma(\boldsymbol{P}|_{\{\sigma,\sigma(\boldsymbol{P})\}})=\sigma(\boldsymbol{P})\neq\sigma$ for all compatible profiles $\boldsymbol{P}\in\mathcal{A}(T,P)$. Hence, by the \hyperref[b]{binary self-selectivity} axiom, there exists some compatible profile $\boldsymbol{P}'\in\mathcal{A}(\{\sigma,\sigma(\boldsymbol{P})\},P)$ such that $\sigma(\boldsymbol{P}')=\sigma\neq\sigma(\boldsymbol{P})$. Thus, $\boldsymbol{P}'\neq\boldsymbol{P}|_{\{\sigma,\sigma(\boldsymbol{P})\}}$. And further, $\sigma(P)=(\sigma(\boldsymbol{P}))(P)$. Now, let $\mu:T\to T$ satisfy $\mu(\sigma)=\sigma(\boldsymbol{P})$, $\mu(\sigma(\boldsymbol{P}))=\sigma$ and $\mu(\sigma')=\sigma'$ for all voting rules $\sigma'\in T\backslash\{\sigma,\sigma(\boldsymbol{P})\}$. By the \hyperref[n]{neutrality} axiom, $\sigma(\mu\boldsymbol{P})=\mu(\sigma(\boldsymbol{P}))=\sigma$. And moreover, $\mu\boldsymbol{P}\in\mathcal{A}(T,P)$. But this contradicts the assumption that there exists no compatible profile $\boldsymbol{P}\in\mathcal{A}(T,P)$ satisfying $\sigma(\boldsymbol{P})=\sigma$. Therefore, $\Sigma_n\cap\Sigma_b\subseteq\Sigma_s$.
\end{proof}

\begin{corollary}\label{col1}
    Under the unrestricted strict preference domain, a \hyperref[u]{unanimous} and \hyperref[n]{neutral} voting rule is \hyperref[b]{binary self-selective} if and only if it is \hyperref[d]{dictatorial}. Formally,
\begin{gather}
    [(\forall\tau\in\mathbb{N})(\mathcal{A}(A_\tau)=\mathcal{P}(A_\tau))]\Rightarrow[(\sigma\in\Sigma_u\cap\Sigma_n)\Rightarrow((\sigma\in\Delta)\iff(\sigma\in\Sigma_b))]
\end{gather}
\end{corollary}

\begin{cproof}
    Let $\mathcal{A}(A_\tau)=\mathcal{P}(A_\tau)$ for all $\tau\in\mathbb{N}$. Then, by \textcite[Theorem 2, p. 990]{koray_00}, a \hyperref[u]{unanimous} and \hyperref[n]{neutral} voting rule $\sigma\in\Sigma_u\cap\Sigma_n$ satisfies $\sigma\in\Sigma_s$ if and only if $\sigma\in\Delta$. By \Cref{th}, a \hyperref[n]{neutral} voting rule $\sigma\in\Sigma_n$ satisfies $\sigma\in\Sigma_b$ if and only if $\sigma\in\Sigma_s$. Therefore, a \hyperref[u]{unanimous} and \hyperref[n]{neutral} voting rule $\sigma\in\Sigma_u\cap\Sigma_n$ satisfies $\sigma\in\Sigma_b$ if and only if $\sigma\in\Delta$.
\end{cproof}

\Cref{th,col1} show that substituting \hyperref[s]{universal self-selectivity} for \hyperref[b]{binary self-selectivity} is not enough to escape \citeauthor{koray_00}'s (\citeyear{koray_00}, Theorem 2, p. 990) impossibility result. Yet, in \Cref{col2,col3}, we obtain two novel and closely related characterizations of the \hyperref[con]{Condorcet voting rule} on the domain of preference profiles with a strong Condorcet winner.

Given any $\tau\in\mathbb{N}$ and any profile $R\in\mathcal{R}(A_\tau)$, let $N_R(x,y)=\{i\in N\mid xR_iy\}$ be the \emph{set of voters who (weakly) prefer $x$ to $y$}, and let $n_R(x,y)=|N_R(x,y)|$ be \emph{its cardinality}. Given any profile $P\in\mathcal{P}(A_\tau)$, let $C_P=\{x\in A_\tau\mid(\forall y\in A_\tau\backslash\{x\})(n_P(x,y)>n/2)\}$ be the \emph{set of strong Condorcet winners at $P$}. Then, $C_P\in\{\emptyset,c_P\}$ (i.e., $C_P$ is either empty or a singleton). Let $\mathcal{C}(A_\tau)=\{P\in\mathcal{P}(A_\tau)\mid C_P\neq\emptyset\}$ be the \emph{set of strict preference profiles on $A_\tau$ with a strong Condorcet winner}.

\begin{definition}[Condorcet voting rule]\label{con}
    Given any strict preference profile with a strong Condorcet winner, the \emph{Condorcet voting rule} $c$ always selects it. Formally,
\begin{gather}
    (\forall\tau\in\mathbb{N})[(\mathcal{A}(A_\tau)=\mathcal{C}(A_\tau))\wedge((\forall P\in\mathcal{A}(A_\tau))(c(P)=c_P))]
\end{gather}
\end{definition}

\begin{theorem}\label{col2}
    Under the strict preference domain with a strong Condorcet winner; a \hyperref[u]{unanimous}, \hyperref[n]{neutral} and \hyperref[a]{anonymous} voting rule is \hyperref[b]{binary self-selective} if and only if it is the \hyperref[con]{Condorcet voting rule}. Formally,
\begin{gather}
    [(\forall\tau\in\mathbb{N})(\mathcal{A}(A_\tau)=\mathcal{C}(A_\tau))]\Rightarrow[(\sigma\in\Sigma_u\cap\Sigma_n\cap\Sigma_a)\Rightarrow((\sigma=c)\iff(\sigma\in\Sigma_b))]
\end{gather}
\end{theorem}

\begin{proof}
    Let $\mathcal{A}(A_\tau)=\mathcal{C}(A_\tau)$ for all $\tau\in\mathbb{N}$. Then, there are three statements to show:
\begin{enumerate}
    \item $c\in\Sigma_u\cap\Sigma_n\cap\Sigma_a$,
    \item $(\sigma\in\Sigma_u\cap\Sigma_n\cap\Sigma_a)\Rightarrow((\sigma=c)\Rightarrow(\sigma\in\Sigma_b))$,
    \item $(\sigma\in\Sigma_u\cap\Sigma_n\cap\Sigma_a)\Rightarrow((\sigma\in\Sigma_b)\Rightarrow(\sigma=c))$.
\end{enumerate}

\begin{statement}\label{s1'}
    $c\in\Sigma_u\cap\Sigma_n\cap\Sigma_a$.
\end{statement}

    The proof is direct. Fix any $\tau\in\mathbb{N}$. Consider any profile $P\in\mathcal{C}(A_\tau)$ satisfying $t_i(P)=x$ for all voters $i\in N$. Then, $c(P)=c_P=x$. Hence, $c\in\Sigma_u$. Consider any profile $P\in\mathcal{C}(A_\tau)$ and any permutation $\mu\in M_\tau$. Then, $\mu(c(P))=\mu(c_P)=c_{\mu P}=c(\mu P)$. Hence, $c\in\Sigma_n$. Consider any profile $P\in\mathcal{C}(A_\tau)$ and any permutation $\pi\in\Pi$. Then, $c(P)=c_P=c_{\pi P}=c(\pi P)$. Hence, $c\in\Sigma_a$. Therefore, $c\in\Sigma_u\cap\Sigma_n\cap\Sigma_a$.

\begin{statement}\label{s2'}
    $(\sigma\in\Sigma_u\cap\Sigma_n\cap\Sigma_a)\Rightarrow((\sigma=c)\Rightarrow(\sigma\in\Sigma_b))$.
\end{statement}

    The proof is direct. By \Cref{s1'}, $c\in\Sigma_u\cap\Sigma_n\cap\Sigma_a$. Thus, we only need to show that $c\in\Sigma_b$. Notice that if $A_2=\{x,y\}$; then, for all profiles $P\in\mathcal{C}(A_2)$, we have $c(P)=x$ if and only if $n_P(x,y)>n/2$. Consider any $\tau\in\mathbb{N}$, any profile $P\in\mathcal{C}(A_\tau)$, and any \hyperref[n]{neutral} voting rule $\sigma\in\Sigma_n\backslash\{c\}$. Fix $T=\{c,\sigma\}$. Then, $n_{P}(c(P),\sigma(P))>n/2$. Thus, $n_{\boldsymbol{R}^{T\!P}}(c,\sigma)>n/2$. Therefore, there exists some compatible profile $\boldsymbol{P}\in\mathcal{C}(T,P)$ such that $n_{\boldsymbol{P}}(c,\sigma)>n/2$. Then, $c(\boldsymbol{P})=c$. Thus, $c\in\Sigma_b$. 

\begin{statement}\label{s3'}
    $(\sigma\in\Sigma_u\cap\Sigma_n\cap\Sigma_a)\Rightarrow((\sigma\in\Sigma_b)\Rightarrow(\sigma=c))$.
\end{statement}

    We now introduce one new axiom that will be useful to prove \Cref{s3'}.

\begin{axiom}[Pareto]\label{p}
    A voting rule $\sigma$ is \emph{Paretian} if and only if it never selects strictly Pareto dominated alternatives. Formally, if and only if
\begin{gather}\label{eq.p}
    (\forall\tau\in\mathbb{N})(\forall P\in\mathcal{A}(A_\tau))(\forall x,y\in A_\tau)[((\forall i\in N)(xP_iy))\Rightarrow(\sigma(P)\neq y)]
\end{gather}    
\end{axiom}

    Let $\Sigma_p=\{\sigma\in\Sigma\mid\sigma\text{ is \hyperref[p]{Paretian}}\}$.
    
    Let $A_2=\{x,y\}$ and let $\Sigma_2=\{\sigma\in\Sigma\mid(\forall P\in\mathcal{C}(A_2))[(\sigma(P)=x)\iff(n_P(x,y)>n/2)]\}$ be the \emph{set of voting rules that select the alternative that is top-ranked by the largest number of voters whenever there are just two alternatives}. By \Cref{s1'}, $c\in\Sigma_u\cap\Sigma_n\cap\Sigma_a$. Thus, the proof of \Cref{s3'} follows directly from these four claims:
\begin{enumerate}[label=3.\arabic*.]
    \item[3.1.] $\Sigma_n\cap\Sigma_b\subseteq\Sigma_i$,
    \item[3.2.] $\Sigma_u\cap\Sigma_i\subseteq\Sigma_p$,
    \item[3.3.] $\Sigma_n\cap\Sigma_a\cap\Sigma_i\cap\Sigma_p\subseteq\Sigma_2$,
    \item[3.4.] $\Sigma_2\cap\Sigma_b=\{c\}$.
\end{enumerate}

\begin{claim}\label{c31'}
    $\Sigma_n\cap\Sigma_b\subseteq\Sigma_i$.
\end{claim}

    The proof follows directly from \Cref{c21} in the proof of \Cref{th}.

\begin{claim}\label{c32'}
    $\Sigma_u\cap\Sigma_i\subseteq\Sigma_p$.
\end{claim}

    The proof is by contradiction. Consider any $\tau\in\mathbb{N}$, any profile $P\in\mathcal{C}(A_\tau)$ and any \hyperref[u]{unanimous} and \hyperref[iia]{independent of irrelevant alternatives} voting rule $\sigma\in\Sigma_u\cap\Sigma_i$ satisfying $\sigma(P)=x$. Let alternative $y\in A_\tau\backslash\{x\}$ satisfy $yP_ix$ for all voters $i\in N$. Then, the \hyperref[iia]{independence of irrelevant alternatives} axiom implies $\sigma(P|_{\{x,y\}})=x$; whereas the \hyperref[u]{unanimity} axiom implies $\sigma(P|_{\{x,y\}})=y$, which is a contradiction. Thus, $\Sigma_u\cap\Sigma_i\subseteq\Sigma_p$.

\begin{claim}\label{c33'}
    $\Sigma_n\cap\Sigma_a\cap\Sigma_i\cap\Sigma_p\subseteq\Sigma_2$.
\end{claim}

    The proof is by contradiction. Consider any \hyperref[n]{neutral}, \hyperref[a]{anonymous}, \hyperref[iia]{independent of irrelevant alternatives} and \hyperref[p]{Paretian} voting rule $\sigma\in(\Sigma_n\cap\Sigma_a\cap\Sigma_i\cap\Sigma_p)\backslash\Sigma_2$. Then, there exists some profile $P\in\mathcal{C}(A_2)$ satisfying $\sigma(P)=x$ and $n_P(x,y)<n_P(y,x).$ Fix any such profile $P\in\mathcal{C}(A_2)$ and consider any coalition $T\subseteq N$ satisfying $|T|=n_P(y,x)$ as well as $yP_ix$ for all voters $i\in N\backslash T$. Then, let $A_3=\{x,y,z\}$ and consider the profile $P'\in\mathcal{C}(A_3)$ satisfying $P'|_{\{x,y\}}=P$; $yP'_iz$ for all voters $i\in N$; $xP'_iz$ for all voters $i\in T$; and $zP'_ix$ for all voters $i\in N\backslash T$. Since $P'|_{\{x,y\}}=P$, it follows that $\sigma(P'|_{\{x,y\}})=\sigma(P)=x$. By the \hyperref[p]{Pareto} axiom, $\sigma(P')\neq z$. By the \hyperref[iia]{independence of irrelevant alternatives} axiom, $\sigma(P')=\sigma(P'|_{\{x,y\}})=x=\sigma(P'|_{\{x,z\}})$. Let $\beta:\{x,z\}\to\{x,y\}$ satisfy $\beta(x)=y$ and $\beta(z)=x$. By the \hyperref[n]{neutrality} axiom, $\sigma(\beta P'|_{\{x,z\}})=\beta(\sigma(P'|_{\{x,z\}}))=\beta(x)=y$. Further, there exists some permutation $\pi\in\Pi$ such that $\pi\beta P'|_{\{x,z\}}=P$. Fix such a permutation $\pi\in\Pi$. Then, $\sigma(\pi\beta P'|_{\{x,z\}})=\sigma(P)=x$. But by the \hyperref[a]{anonymity} axiom, $\sigma(\pi\beta P'|_{\{x,z\}})=\sigma(\beta P'|_{\{x,z\}})=y$, which is a contradiction. Thus, $\Sigma_n\cap\Sigma_a\cap\Sigma_i\cap\Sigma_p\subseteq\Sigma_2$.

\begin{claim}\label{c34'}
    $\Sigma_2\cap\Sigma_b=\{c\}$.
\end{claim}

    The proof is by contradiction. Consider any \hyperref[b]{binary self-selective} voting rule $\sigma\in(\Sigma_2\cap\Sigma_b)\backslash\{c\}$. Then, there exists some $\tau\in\mathbb{N}$ and some profile $P\in\mathcal{C}(A_\tau)$ such that $\sigma(P)\neq c_P$. Fix any such profile $P\in\mathcal{C}(A_\tau)$. Let $T=\{c,\sigma\}$. Then, $n_P(c(P),\sigma(P))>n/2$. Thus, $\mathcal{C}(T,P)=\{\boldsymbol{P}\}$, where $n_{\boldsymbol{P}}(c,\sigma)>n/2$. Since $\sigma\in\Sigma_2$, it follows that $\sigma(\boldsymbol{P})=c$, thus contradicting the assumption that $\sigma\in\Sigma_b$. Hence, $\sigma\notin\Sigma_b$. Therefore, $\Sigma_2\cap\Sigma_b=\{c\}$.
\end{proof}

\begin{corollary}\label{col3}
    Under the strict preference domain with a strong Condorcet winner; a \hyperref[u]{unanimous}, \hyperref[n]{neutral} and \hyperref[a]{anonymous} voting rule is \hyperref[s]{universally self-selective} if and only if it is the \hyperref[con]{Condorcet voting rule}. Formally,
\begin{gather}
    [(\forall\tau\in\mathbb{N})(\mathcal{A}(A_\tau)=\mathcal{C}(A_\tau))]\Rightarrow[(\sigma\in\Sigma_u\cap\Sigma_n\cap\Sigma_a)\Rightarrow((\sigma=c)\iff(\sigma\in\Sigma_s))]
\end{gather}
\end{corollary}

\begin{cproof}
    The proof is direct. Let $\mathcal{A}(A_\tau)=\mathcal{C}(A_\tau)$ for all $\tau\in\mathbb{N}$. By \cref{col2}, a \hyperref[u]{unanimous}, \hyperref[n]{neutral} and \hyperref[a]{anonymous} voting rule $\sigma\in\Sigma_u\cap\Sigma_n\cap\Sigma_a$ satisfies $\sigma\in\Sigma_b$ if and only if $\sigma=c$. By \Cref{th}, a \hyperref[n]{neutral} voting rule $\sigma\in\Sigma_n$ satisfies $\sigma\in\Sigma_b$ if and only if $\sigma\in\Sigma_s$. Therefore, a \hyperref[u]{unanimous}, \hyperref[n]{neutral} and \hyperref[a]{anonymous} voting rule $\sigma\in\Sigma_u\cap\Sigma_n\cap\Sigma_a$ satisfies $\sigma\in\Sigma_s$ if and only if $\sigma=c$.
\end{cproof}

\acknowledgments{We are very thankful for their helpful comments and suggestions to Marc Claveria-Mayol, Andrea Marietta Leina, Ben McQuillin, R\'{o}bert Somogyi, Christopher Stapenhurst, Anastas P. Tenev, QSMS members, the Associate Editor and three anonymous referees. Also, Toygar T. Kerman gratefully acknowledges funding by the Hungarian National Research, Development and Innovation Office, Project Number K-143276. No generative AI was used in the write-up of this paper. All errors are only ours. We declare no conflict of interest.}

\conflictofinterest{We declare no conflict of interest.}

\dataavailability{No data was used to write this paper.}

\funding{None.}

\printbibliography[]
\end{document}